\documentclass[a4paper,20pt]{article}
    \usepackage[T1]{fontenc}
    \usepackage{graphicx}
    \usepackage{epsfig}
    \usepackage{amsmath}
    \usepackage{amsfonts}
    \renewcommand{\abstract}{}
\begin{document}
\makeatletter
\renewcommand{\@oddhead}{\textit{Advances in Astronomy and Space Physics} \hfil \textit{V.A. Sadova, A.V. Tugay}}
\renewcommand{\@evenfoot}{\hfil \thepage \hfil}
\renewcommand{\@oddfoot}{\hfil \thepage \hfil}
\fontsize{11}{11} \selectfont

\title{Luminosity - spectral index dependence of X-ray bright Seyfert galaxies }
\author{\textsl{V.A. Sadova, A.\,V.~Tugay}}
\date{}
\maketitle
\begin{center} {\small 
Taras Shevchenko National University of Kyiv, Glushkova ave., 4, 03127, Kyiv, Ukraine\\
tugay@anatoliy@gmail.com}
\end{center}

\begin{abstract}

X-ray luminosities and spectral indices of 97 bright Seyfert 1 galaxies from XMM archive are analyzed in this article. Distribution of these values is random so we conclude that the model of emission should be at least two-parametric. Within the merging model of AGN the relation between black hole mass, stage of merging and observable X-ray parameters is proposed.

\end{abstract}

\section*{Introduction}
\indent \indent XMM-Newton observations archive is the largest and the most convenient database for the analysis of X-ray spectra of any celestial bodies including extragalactic.
Xgal sample of X-ray galaxies (\cite{tugay12}, \cite{tugay14a}) contains more than 4000 XMM sources associated with galaxies or galaxy clusters. The main goal of compiling Xgal was the study of a large-scale structure (LSS) of the Universe in X-ray band (0.2-15 keV for XMM). The distribution of the main elements of LSS - filaments, voids and walls - can be recovered for redshifts up to 0.2 (\cite{tugay13}, \cite{tugay14}). It was shown in \cite{tugay11} and \cite{tugay13a} that the most frequent type of X-ray emitting galaxies at such distances is Seyfert 1.
Spectra of Seyfert 1 galaxies in SDSS region were analyzed in \cite{tugay13a}.
30 of them are Compton thin ($N_H<10^{25} cm^{-2}$) and thus can be correctly fitted by the powerlaw. 
The rest bright Sy 1 X-ray galaxies with radial velocities from 4000 to 39000 km/s are studied in present work. We compiled a list of major X-ray galaxies in nearby Universe - Compton thin Sy 1 galaxies. Our goal was to calculate spectral parameters of these galaxies, to  build the distribution of that parameters and to choose some realistic connection between observed spectral parameters and the parameters of internal structure of AGN.  

\begin{figure}[!h]
\begin{minipage}[t]{.99\linewidth}
\centering
\epsfig{file = 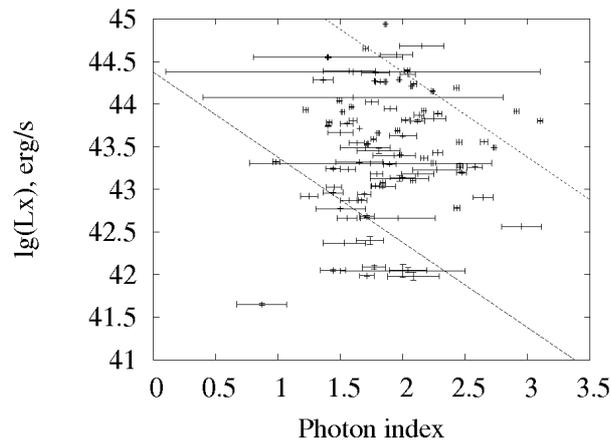, width = 0.49\linewidth}
\caption{X-ray parameters of Seyfert 1 galaxies. Lower line marks the mass of the central black hole $10^7 M_\odot$; for upper line $M_{BH}=10^9 M_\odot$}\label{fig1}
\end{minipage}
\end{figure}

\section*{Sample selection and X-ray spectral analysis}
\indent \indent The statistics of galaxies considered for this work is the following.
There are 582 bright X-ray extragalactic sources outside SDSS region. 
87 of them are Seyfert 1. The resulting list of bright Sy 1 galaxies from Xgal sample consists of three parts:

1. Spectral parameters for 30 Sy 1 galaxies in SDSS region were obtained previously in \cite{tugay13a}.

2. 23 new spectra were built in the present work using the standard XMM SAS package. We get event lists for PN camera with epproc procedure, filtered them from solar protons with parameters 150<PI<15000 and PATTERN=0 and derived spectra with especget procedure. Background region was selected from the same CCD chip as the source and of the same size.

3. 46 spectra were found from literature. 33 of them were from CAIXA - Catalog of AGNs in XMM Archive (\cite{bianchi09}).

Parameters of Sy 1 galaxies from p.2 and p.3, except for CAIXA entries, are presented in Tables 1 and 2. 
The main observable parameters of X-ray emission are luminosity and a spectral index. In most cases of compton thin Sy 1 galaxies these are the only X-ray emission parameters that can be fitted correctly. The distribution of these parameters is presented at Fig. 1. 
Another important spectral feature in 2-15 keV energy band is the iron emission line at 6.5 keV, but it is detected rarely, so we do not consider it here. Also we find two galaxies with thermal dominated emission: 2E1891 and IRAS05218-1212. We found the best fit blackbody temperature $1.11\pm 0.03 $ keV and $2.37\pm 0.13 $ keV for these galaxies respectively. The powerlaw component was not fitted correctly for these galaxies so they were excluded from further analysis.

\section*{Interpretation of X-ray luminosity and spectral index}
\indent \indent Previously analyzed parameters should be connected with some intrinsic parameters of AGN.
In the model of Hopkins et al. (2008, \cite{hopkins08}) AGN appears as a result of merging of two galaxies. Different observable features of AGN can be interpreted as stages of merging.
According to this model it was supposed the simplest relation that the photon index equals to a decimal logarithm of time since collision:

\begin{equation}\label{form1}
\Gamma=lg(t / 10^6 years).
\end{equation}

It was assumed here that at large times after merging hard X-ray emission decreases that could appear as the increasing spectral index. Since the central engine of AGN is assumed to be a supermassive black hole in its center, luminosity of AGN should correlate with the black hole mass. Taking into account the decreasing luminosity with the age, the following formula for estimating of the black hole mass is proposed: 

\begin{equation}\label{form2}
lg(M_{BH}/M_\odot )=lgL_X+\Gamma-A,
\end{equation}

\noindent where $L_X$ is measured in erg/s. The physical reason of (2) is the assumption that the total amount of the emitted energy ($L_X \cdot t$) should be related to the total energy budget of the source (mass of the available gas, which might be proportional to $M_{BH}$). This relation has not yet been verified and may have unclear systematical uncertainty behind.
Coefficient A=37.375 was selected to equalize the average black hole mass with the results of a similar work of Vestergaard and Peterson (2006, \cite{vestergaard06}, hereafter VP), where the black hole masses for AGNs were also estimated from X-ray emission. The averaged logarithm of the black hole mass in \cite{vestergaard06} and for our sample  equals (at 1$\sigma$ level)

\begin{equation}\label{form3}
lg(M_{BH}/M_\odot )=7.992\pm 0.562.
\end{equation}

Two lines corresponding to the black hole mass of $10^7 M_\odot$ and $10^9 M_\odot$ are shown at Fig. 1. Masses of the central black holes of individual galaxies are presented in the last column of Table 2. The uncertainty of $lg(M_{BH}/M_\odot )$ depends of the uncertainties of $\Gamma$, $L_X$ and systematic effects. According to (3), this uncertainty should be approximately equal to 0.5 or less.
Black hole mass distributions for our sample and VP galaxies were approximated by Gaussian. Deviation for Xgal Sy1 galaxies appears somewhat larger than in \cite{vestergaard06} - $\sigma=0.826$. 
This value is potentially biased, as here we study solely the X-ray bright objects, whereas the VP sample, based on optical data, include both X-ray bright and dim sources.
The distributions of the black hole masses for our sample and VP results are presented in Table 3. The percentage of normal Gaussian distribution is also given for the comparison.
Width of bin is equal to 1 $\sigma $. 

\section*{Conclusion and discussion}
\indent \indent 

Distribution of X-ray emission parameters at Fig. 1 is random so we conclude that the model of galaxy emission should be at least two-parametric. 
We propose to consider the black hole mass and the merging stage as such intrinsic model parameters.
Dependence of X-ray spectral index from X-ray luminosity was recently found in \cite{yang15}. The authors considered the dependence of $\Gamma $ from $L_X/L_{Edd}$ ratio and interpreted this dependence by two-phase advection dominated accretion. In such study
$L_{Edd}$ and black hole mass should be estimated independent from luminosity (for the method of $M_{BH}$ estimation by AGN X-ray emission see, for example, \cite{jang14} ). It is possible in the case if the data of X-ray variability is available and the same source can show different values of $\Gamma $ and $L_X/L_{Edd}$ in the series of observations. The most Xgal object has only one XMM observation and all X-ray lightcurves for our Seyfert 1 galaxies are constant. So we can not consider $\Gamma (L_X/L_{Edd})$ dependence and conclude here the parameter no significant dependence of  $\Gamma $ from $L_X$.

\begin{table}
 \centering
 \caption{General parameters of new Seyfert 1 galaxies added to analysis. The rest sample see in \cite{tugay13a}. u - u-band apparent magnitude; r - major semiaxis of 25$^m/''$ contour; V$_3K$ - radial velocity in CMB reference frame. Parameters were taken from Hyperleda database. }\label{tab1}
 \vspace*{1ex}
 \begin{tabular}{rrrrrrrr}
  \hline
N & Name & RA, deg & DEC, deg & Coord. num. & u & r, arcsec & V$_{3K}$, km/s   \\
\hline
 1 & 2MASX J00044124+0007113  &   1,1718 &   0,1198 &  0004+0007  & 18,63 &  6,3 & 31923 \\  
 2 &               ESO 540-1  &   8,5571 & -21,4389 &  0034-2126  & 13,71 & 37,8 &  7744 \\  
 3 & 2MASX J00440466+0101531  &  11,0195 &   1.0313 &  0044+0101  & 17,77 &  9,3 & 33210 \\  
 4 & 2MASX J00565517-7513524  &  14,2297 & -75,2312 &  0056-7513  & 15,04 &  6,4 & 22137 \\  
 5 &                 Mrk 993  &  21,3812 &  32,1360 &  0125+3208  & 14,37 & 47,6 &  4380 \\  
 6 &                   3C 59  &  31,7590 &  29,5128 &  0207+2930  & 17,44 & 11,1 & 32612 \\  
 7 &                UGC 1841  &  35,7989 &  42,9914 &  0223+4259  & 13,75 & 65,7 &  6167 \\  
 8 & 2MASX J02491286-0815254  &  42,3036 &  -8,2571 &  0249-0815  & 16,64 & 14,0 &  8617 \\  
 9 &              ESO 359-19  &  61,2570 & -37,1876 &  0405-3711  & 15,52 & 17,3 & 16476 \\  
10 &                  3C 111  &  64,5885 &  38,0266 &  0418+3801  & 19,75 & 10,0 & 15305 \\  
11 &                Mrk 1506  &  68,2962 &   5,3542 &  0433+0521  & 15,06 & 23,3 &  9839 \\  
12 &               ESO 15-11  &  68,8183 & -78,0323 &  0435-7801  & 15,58 & 20,3 & 18351 \\  
13 &                 RBS 560  &  69,3672 & -47,1916 &  0437-4711  & 16,61 &  7,4 & 15574 \\  
14 &                UGC 3142  &  70,9449 &  28,9718 &  0443+2858  & 15,84 & 28,0 &  6434 \\  
15 &                Pictor A  &  79,9570 & -45,7789 &  0519-4546  & 16,25 & 14,7 & 10516 \\  
16 &         IRAS 05218-1212  &  81,0288 & -12,1693 &  0524-1210  & 15,70 &  7,6 & 14721 \\  
17 &                 2E 1644  &  95,7820 & -64,6060 &  0623-6436  & 17,06 & 15,0 & 36197 \\  
18 & 2MASX J07185777+7059209  & 109,7410 &  70,9891 &  0719+7059  & 17,40 &  5,9 & 19810 \\  
19 &                 2E 1891  & 119,5000 &  39,3414 &  0754+3928  & 15,21 &  2,0 & 28935 \\  
20 &            Sextans Ring  & 150,5010 &  -8,1614 &  0959-0809  & 15,22 & 13,0 &  4910 \\  
21 &          MCG +11-19-030  & 239,2650 &  63,8408 &  1557+6350  & 15,42 & 20,8 &  9000 \\  
22 & 2MASX J16115141-6037549  & 242,9640 & -60,6319 &  1611-6037  & 14,70 & 26,7 &  4777 \\  
23 & 2MASX J16174561+0603530  & 244,4400 &   6,0649 &  1617+0603  & 16,19 & 14,4 & 11479 \\  
24 &                 Mrk 883  & 247,4700 &  24,4439 &  1629+2426  & 15,78 & 18,1 & 11447 \\  
25 &                 2E 4097  & 278,7640 &  32,6964 &  1835+3241  & 15,15 & 21,2 & 17289 \\  
26 &                 FRL 339  & 302,9930 & -57,0868 &  2011-5705  & 16,12 & 14,4 & 16274 \\  
27 &                4C 74.26  & 310,6560 &  75,1341 &  2042+7508  & 15,33 &  2,0 & 31071 \\  
28 &                 Mrk 509  & 311,0410 & -10,7235 &  2044-1043  & 13,35 & 17,7 & 10045 \\  
29 & 2MASX J21022164+1058159  & 315,5900 &  10.9711 &  2102+1058  & 14,92 & 12,6 &  8336 \\  
30 & 2MASX J22191855+1207531  & 334,8270 &  12.1315 &  2219+1207  & 17,19 &  8,3 & 24229 \\  
31 &                  3C 445  & 335,9570 &  -2,1036 &  2223-0206  & 17,26 &  8,9 & 16510 \\  
32 &                NGC 7469  & 345,8150 &   8,8739 &  2303+0852  & 12,90 & 41,4 &  4545 \\  
33 &                NGC 7589  & 349,5650 &   0,2612 &  2318+0015  & 15,23 & 28,7 &  8578 \\  
34 &                NGC 7603  & 349,7360 &   0,2440 &  2318+0014  & 14,04 & 36,1 &  8484 \\  
35 &                NGC 7720  & 354,6230 &  27,0317 &  2338+2701  & 13,43 & 45,4 &  8695 \\  
36 &          MCG -05-01-013  & 359,3665 & -30,4613 &  2357-3027  & 14,96 & 16,9 &  8744 \\ \hline 
 \end{tabular}

\end{table}

\begin{table}
 \centering
 \caption{X-ray parameters of Seyfert 1 galaxies. $F_X$ - X-ray flux in 2-10 keV band in $10^{-14} erg/s/cm^{-2}$; $L_{X40}$ - X-ray luminosity in redshift space for H=70 km/s/Mpc divided by $10^{40}$ erg/s; $\Gamma$ - spectral index; $N_H$ - neutral hydrogen column density; for the spectra fitted in this work $\chi ^2/dof$ is presented instead of reference.}\label{tab2}
 \vspace*{1ex}
 \begin{tabular}{rrrrrrrccr}
\hline
N & Target & $F_X$ & $\Delta F_X$ & $L_{X40}$ & $\Gamma$ & $\Delta \Gamma$ & $N_H, 10^{20}cm^{-2}$ & Ref & $lg(M_{BH}/M_\odot )$  \\
\hline
 1 & 0004+0007 &   45,6 &  1,9 &  1080 & 1,840 & 0,098 & 0,013$\pm $0,011   & 28,46/29            & 7,873 $\pm$0,233\\ 
 2 & 0034-2126 &   80,1 &  2,2 &   112 & 1,440 & 0,100 &                    & \cite{gallo06}      & 6,489 $\pm$0,235\\ 
 3 & 0044+0101 &  110,5 &  8,7 &  2831 & 1,806 & 0,174 & 0,063$\pm $0,027   & 74,98/34            & 8,258 $\pm$0,309\\ 
 4 & 0056-7513 &  557,0 & 14,6 &  6339 & 2,118 & 0,056 & 0,046$\pm $0,008   & 113,13/61           & 8,920 $\pm$0,191\\ 
 5 & 0125+3208 &  216,6 &  3,4 &    96 & 1,710 & 0,060 & 0,07$\pm $0,01     & \cite{corral05}     & 6,694 $\pm$0,195\\ 
 6 & 0207+2930 & 1434,4 &  4,5 & 35428 & 1,398 & 0,006 & 0,0013$\pm $0,0007 & 4765,6/998          & 8,947 $\pm$0,141\\ 
 7 & 0223+4259 &   50,4 &  1,4 &    45 & 0,870 & 0,200 &                    & \cite{croston03}    & 5,518 $\pm$0,335\\ 
 8 & 0249-0815 &   65,3 &  4,2 &   113 & 2,041 & 0,151 & 0,013$\pm $0,015   & 18,07/23            & 7,094 $\pm$0,286\\ 
 9 & 0405-3711 &  612,4 &  7,9 &  3861 & 1,764 & 0,019 & 0$\pm $0,006       & 28,12/17            & 8,351 $\pm$0,154\\ 
10 & 0418+3801 & 8202,2 & 17,5 & 44619 & 1,700 & 0,020 & 0,8                & \cite{lewis05}      & 9,349 $\pm$0,155\\ 
11 & 0433+0521 & 8159,1 & 16,0 & 18343 & 1,860 & 0,010 & 0,01$\pm $0,001    & \cite{ballantyne04} & 9,122 $\pm$0,145\\ 
12 & 0435-7801 &  251,9 &  4,2 &  1970 & 1,894 & 0,050 & 0.051$\pm $0,008   & 78,48/65            & 8,188 $\pm$0,185\\ 
13 & 0437-4711 & 1203,9 &  5,6 &  6781 & 2,247 & 0,097 & 0$\pm $0,001       & 598,7/248           & 9,078 $\pm$0,232\\ 
14 & 0443+2858 & 2179,5 & 16,2 &  2095 & 0,985 & 0,029 & 1,217$\pm $0,045   & 498.34/211          & 7,307 $\pm$0,164\\ 
15 & 0519-4546 & 1784,9 &  4,7 &  4584 & 1,800 & 0,010 & 0,03$\pm $0,01     & \cite{vasylenko15}  & 8,461 $\pm$0,145\\ 
16 & 0524-1210 &  454,4 &  8,6 &  2287 & 9,500 & 9,900 & 0,909$\pm $0,464   & 143,54/28           & - \\ 
17 & 0623-6436 &  818,0 &  8,4 & 24890 & 2,034 & 0,025 & 0,021$\pm $0,003   & 246,18/213          & 9,430 $\pm$0,160\\ 
18 & 0719+7059 &  147,5 & 12,3 &  1344 & 1,971 & 0,235 & 0,024$\pm $0,045   & 2,15/5              & 8,098 $\pm$0,490\\ 
19 & 0754+3928 &  382,1 &  5,0 &  7429 & 7,372 & 0,621 & 0,234$\pm $0,043   & 396,04/40           & - \\ 
20 & 0959-0809 & 1078,5 &  6,7 &   604 & 2,432 & 0,023 & 0,066$\pm $0,002   & 783.18/540          & 8,213 $\pm$0,158\\ 
21 & 1557+6350 &   50,7 &  6,0 &    95 & 2,082 & 0,206 & 0$\pm $0,072       & 5,77/5              & 7,061 $\pm$0,476\\ 
22 & 1611-6037 &  907,8 & 27,7 &   481 & 1,712 & 0,059 & 0,131$\pm $0,016   & 67,21/56            & 7,394 $\pm$0,194\\ 
23 & 1617+0603 & 1596,9 & 14,6 &  4887 & 1,957 & 0,017 & 0,021$\pm $0,002   & 535,78/495          & 8,646 $\pm$0,152\\ 
24 & 1629+2426 &  287,9 &  4,2 &   876 & 1,689 & 0,054 & 0,103$\pm $0,014   & 53,73/57            & 7,632 $\pm$0,189\\ 
25 & 1835+3241 & 6930,5 & 24,1 & 48110 & 2,150 & 0,180 & 3$\pm $1           & \cite{torresi10}    & 9,832 $\pm$0,315\\ 
26 & 2011-5705 &  296,9 &  3,8 &  1826 & 2,576 & 0,062 & 0,039$\pm $0,006   & 94,80/67            & 8,838 $\pm$0,197\\ 
27 & 2042+7508 & 3857,2 & 10,0 & 86479 & 1,860 & 0,010 & 0,183              & \cite{ballantyne05} & 9,797 $\pm$0,145\\ 
28 & 2044-1043 & 8290,5 & 10,2 & 19427 & 1,970 & 0,010 & 0,27$\pm $0,01     & \cite{vasylenko15}  & 9,258 $\pm$0,145\\ 
29 & 2102+1058 &  154,8 & 17,6 &   250 & 1,736 & 0,108 & 0,131$\pm $0,025   & 28,12/17            & 7,134 $\pm$0,376\\ 
30 & 2219+1207 &  466,2 &  4,1 &  6356 & 3,099 & 0,019 & 0,074$\pm $0,002   & 523,61/233          & 9,902 $\pm$0,154\\ 
31 & 2223-0206 &  875,2 & 12,0 &  5540 & 1,4   & 0,1   &  1 - 10            & \cite{sambruna07}   & 8,143 $\pm$0,235\\ 
32 & 2303+0852 & 5311,5 & 14,5 &  2548 & 1,980 & 0,010 & 45$\pm $2          & \cite{vasylenko15}  & 8,387 $\pm$0,145\\ 
33 & 2318+0015 &   71,2 &  2,7 &   122 & 1,768 & 0,093 & 0,036$\pm $0,021   & 11,84/21            & 6,854 $\pm$0,228\\ 
34 & 2318+0014 & 4597,2 & 32,9 &  7685 & 2,280 & 0,030 & 14,8$\pm $5,6      & \cite{singh11}      & 9,165 $\pm$0,165\\ 
35 & 2338+2701 &   62,9 & 12,1 &   110 & 2     & 0,5   & 0,5$\pm $0,2       & \cite{hardcastle05} & 7,041 $\pm$0,905\\ 
36 & 2357-3027 & 1060,8 & 46,5 &  1884 & 2,455 & 0,028 & 0,096$\pm $0.002   & 401,5/302           & 8,729 $\pm$0,163\\ \hline
 \end{tabular}

\end{table}

\begin{table}
 \centering
 \caption{Distribution of black hole mass in units of $\sigma$.  }\label{tab3}
 \vspace*{1ex}
 \begin{tabular}{ccccccccc}
\hline
Interval                                    & <-3 & (-3,-2) & (-2,-1) & (-1,0) & (0,1) & (1,2) & (2,3) & >3 \\
\hline
 Number of galaxies                         &   1 &   2 &   14 &   28 &   37 &   15 &   0 &   0 \\
 \%                                         & 1,0 & 2,1 & 14,4 & 28,9 & 38,1 & 15,5 &   0 &   0 \\
 Number of galaxies in \cite{vestergaard06} &   1 &   0 &    1 &   12 &   13 &    5 &   0 &   0 \\
 \%                                         & 3,1 &   0 &  3,1 & 37,5 & 40,6 & 15,6 &   0 &   0 \\
 Normal distribution, \%                    & 0,2 & 2,1 & 15,6 & 32,1 & 32,1 & 15,6 & 2,1 & 0,2 \\ \hline
 \end{tabular}

\end{table}

\section*{Acknowledgement}
\indent \indent This work was performed at VIRGO.UA. The authors are thankful to ISDC High-Energy Astrophysics data centre for developing Ukrainian Virtual Roentgen and Gamma Observatory. A.Tugay thanks to anonymous referee for useful comments.


\begin{thebibliography}{25}
{\small
\bibitem{ballantyne04} Ballantyne~D.R., Fabian~A.C., Iwasawa~K. 2004, MNRAS, 354, 839  
\bibitem{ballantyne05} Ballantyne~D.R. 2005, MNRAS, 362, 1183  
\bibitem{bianchi09} Bianchi S., Bonilla N.F., Guainazzi M., Matt G., Ponti G. 2009, A$\& $A, 501, 915  
\bibitem{corral05} Corral~A., Barcons~X., Carrera~F.J. et al. 2005, A$\& $A, 431, 97  
\bibitem{croston03} Croston~J.H., Hardcastle~M.J., Birkinshaw~M., Worrall~D.M. 2003, MNRAS, 346, 1041  
\bibitem{gallo06} Gallo~L.C., Lehmann~I., Pietsch~W. et al. 2006, MNRAS, 365, 688  
\bibitem{hardcastle05} Hardcastle~M.J., Sakelliou~I., Worrall~D.M. 2005, MNRAS, 359, 1007  
\bibitem{hopkins08} Hopkins~P.F., Hernquist~L., Cox~T.J. and Keres~D. 2008, ApJS, 175, 356 
\bibitem{jang14} Jang I., Gliazzi M., Hughes C., Titarchuk L. 2014, MNRAS, 443, 72
\bibitem{lewis05} Lewis~K.T., Eracleous~M., Gliozzi~M. 2005, ApJ, 622, 816  
\bibitem{sambruna07} Sambruna~R.M., Reeves~J.N., Braito~V. 2007, ApJ, 665, 1030  
\bibitem{singh11} Singh~V., Shastri~P., Risaliti~G. 2011, A$\& $A, 532, 84  
\bibitem{torresi10} Torresi~E., Grandi~P., Longinotti~A.L. et al. 2010, MNRAS, 401, 10  
\bibitem{tugay11} Tugay~A.\,V. \& Vasilenko~A.\,A. 2011, Odessa Astronomical Publications, 24, 72
\bibitem{tugay12} Tugay~A.\,V. 2012, Odessa Astronomical Publications, 25, 142
\bibitem{tugay13} Tugay~A.\,V. 2013, AASP, 3, 116
\bibitem{tugay13a} Tugay~A.\,V. 2013, Astronomical School's Report, 9, 64
\bibitem{tugay14} Tugay~A.\,V. 2014, AASP, 4, 42
\bibitem{tugay14a} Tugay~A.\,V. 2014, Proceedings of the IAU, 304, 168
\bibitem{vasylenko15} Vasylenko~A., Zhdanov~V., Fedorova~E. 2015, Ap$\& $SS, in press, [arXiv:1501.04555]  
\bibitem{vestergaard06} Vestergaard~M., Peterson~B.M. 2006, ApJ, 641, 689 
\bibitem{yang15} Yang Q., Xie F., Yuan F. et al. 2015, MNRAS, 447, 1692

}
\end{thebibliography}
\end{document}